# Active Gate Drive with Gate-Drain Discharge Compensation for Voltage Balancing in Series-Connected SiC MOSFETs

Ye Zhou, Xu Wang, Liang Xian, *Member IEEE*, and Dan Yang

*Abstract-* Imbalanced voltage sharing during the turn-off transient is a challenge for series-connected silicon carbide (SiC) MOSFET application. This article first discusses the influence of the gate-drain discharge deviation on the voltage imbalance ratio, and its primary causes are also presented and verified by LTspice simulation. Accordingly, a novel active gate drive, which aims to compensate the discharge difference between devices connected in series, is proposed and analyzed. By only using the original output of the driving IC, the proposed gate drive is realized by implementing an auxiliary circuit on the existing commercial gate drive. Therefore, unlike other active gate drives for balancing control, no extra isolations for power/signal are needed, and the number of the devices in series is unlimited. The auxiliary circuit includes three sub-circuits as a high-bandwidth current sink for regulating switching performance, a relative low-frequency but reliable sampling and control circuit for closed-loop control, and a trigger combining the former and the latter. The operational principle and the design guideline for each part are presented in detail. Experimental results validate the performance of the proposed gate drive and its voltage balancing control algorithm.

*Index Terms—* SiC MOSFET, series-connected, active gate drive, discharge compensation, voltage balance.

I. INTRODUCTION

Power electronic applications such as high-speed electrified railway traction and high voltage direct current (HVDC) transmission system put forward higher demands for power semiconductor devices with high nominal voltage rating and switching frequency [1]-[7]. Among the commercially available power semiconductors, SiC MOSFET has got wide attention thanks to its superior properties such as high breakdown field strength and high switching speed. However, the current manufacturing craft and cost limit the maximum voltage-blocking capability of commercial SiC MOSFET to less than 1.7kV [2], which imply that a single SiC MOSFET cannot meet the requirements of the aforesaid high voltage applications.

Compared to using multilevel inverter topologies, combining multiple semiconductors in series-connection is an effective solution to enhance the voltage-blocking capability due to more simplified control and less number of components. Furthermore, according to previous research shown in [3], several low-voltage semiconductors in series-connection offer smaller conduction loss, lower cost and shorter switching transient in comparison with a single high-voltage semiconductor. The major constraint for the application of series-connected power devices is unequal voltage distribution across them in both static and dynamic conditions. The static voltage imbalance caused by different dynamic voltage sharing and equivalent off-state resistances can be solved by paralleling balancing resistors. It is a trade-off between voltage imbalance ratio and power consumption when choosing the values of those balancing resistors [4]. Even harder to be solved, the dynamic voltage unbalancing problems caused by mismatches on gate drive circuits, device characteristics and parasitic capacitances have recently drawn a lot of research focuses [4]-[26].

Previously reported work has proposed many schemes for equal dynamic voltage sharing. They can be divided into three categories:

1) Passive snubber circuits [5]-[8]. Implementing a snubber circuit composed of passive elements (capacitors, resistors, diodes, etc.) across each power device can regulate the voltage changing ratio and ensure balanced voltage. However, the incremental power loss generated by the passive components restricts the widespread applications of this method. Even utilizing a concentrated clamp mode snubber proposed in [5] and [7], the loss produced in transistors by prolonged switching transients still cannot be ignored.

2) Capacitive coupling drives [9]-[15]. With the help of coupling capacitors, only one external driving circuit can drive all switches in series-connection. Although simple circuit structure and high switching frequency can be achieved, the voltage imbalance during the turn-off period still exists. In general, balanced voltage sharing by these methods is gained by the calculated coupling capacitors, the selection of which is according to the datasheet of the power semiconductor device [14][15]. Due to the deviation of parameters among series-connected semiconductors, turn-off delay of high-side devices

Manuscript received April 9, 2020; revised May 15, 2020 and July 14, 2020; accepted September 10, 2020. This work was supported in part by the National Natural Science Foundation of China under Grant Nos. 51607029 and 61836011. The authors also thanks for supporting by the Fundamental Research Funds for the Central Universities 2020GFZD008 and 2020-GFYD011.

Y. Zhou and X. Wang are with the School of Information Science and Engineering, Northeastern University, Shenyang 110819, China (e-mail: 308744199@163.com; wangxu@mail.neu.edu.cn).

L. Xian is with the Energy Research Institute @ Nanyang Technological University, 627590, Singapore (e-mail: liang.xian@ntu.edu.sg).

D. Yang is with the Key Laboratory of Infrared Optoelectric Materials and Micro-Nano Devices, Northeastern University, Shenyang 110819, China (e-mail: yangdan@mail.neu.edu.cn).

cannot get fine compensated, especially under the varying temperature or duty cycle circumstances. Moreover, according to the design criteria discussed in [11]-[13], the gate charge and discharge of the upper devices are supplied by the difference between the charges of adjacent coupling capacitors. With the increase of the stacked devices number, in order to ensure a reliable operation of the uppermost device, the value of the lowermost coupling capacitor also needs to be increased, which will slow down the switching speed of the entire stacked semiconductors. This contradiction limits the number of series devices to some extent. Finally, due to the gate-source voltage deviations, it is difficult to avoid uneven switching loss among devices.

3) Active gate control methods [16]-[26]. These methods modify the switching behavior of the semiconductor by controlling the electric quantity of the input capacitance $C_{iss}$ and the time instant when $C_{iss}$ is charged or discharged. Regulating time delay among the drive signals from the controller has been focused owing to the fast development of high-speed microcontroller, especially, field-programmable gate array (FPGA). In [16]-[18], an active voltage clamping circuit in series with a sampling resistor is placed between collector and gate to protect the IGBT from overvoltage and transfer the clamping time back to the control board. Accordingly, the controller can adjust the gate signals based on the activation time of the clamping circuit to achieve voltage balance. However, the extra loss generated by transient voltage suppressor (TVS) diodes and IGBTs operating in active region limits the application of this technique in high-frequency power systems. A drain-to-source voltage feedback circuit composed of dividing resistors is proposed in [19] and [20] for time adjustment. However, this method introduces additional signals, which need auxiliary high-voltage isolation components for transmission. Beyond that, additional modification of algorithm will put forward higher requirements on the memory and operating frequency for the controller of the power converter. Finally, prolonged response time caused by delayed signal will reduce the control precision and the consistency of drive pulse-width [27]. Another option for controlling the gate charge is to regulate the amplitude of the gate voltage or current in real-time. In [21], a gate drive with active voltage control based on the instantaneous state of SiC MOSFET is proposed to adjust the drain-to-source voltage during its rising transient. However, high rising slope and oscillations of the drain-to-source voltage of SiC MOSFET make it hard for the control loop to follow up on time. Active gate current control methods proposed in [22]-[25] have shown good performance on voltage balance, but they are designed only for Si IGBTs due to the long propagation time of status detection circuits or digital-to-analog converters (DAC). Auxiliary gate current generated by an external Miller capacitor in [26] is an innovative and effective way to acquire superior voltage balancing performance. However, the capacitor compensation circuit needs additional isolated power supplies and other high-bandwidth analog parts, which adds the complexity and cost of the drive.

To overcome the above drawbacks, a novel active gate drive consisting of three sub-circuits is proposed in this study. Several advantages over other voltage balancing techniques are stated as follows:

1) The proposed gate drive does not require additional isolation barriers for signal and power supply.

2) The auxiliary circuit is triggered by falling edges only, which means it can be easily integrated with most of the commercial drive ICs through their output voltage pins.

3) The trigger circuit for combining high-response parts with relatively low-frequency control algorithm makes the bandwidth of the sampling and control circuit the same as the switching frequency of the semiconductor. Therefore, low-frequency, low-cost Microcontroller Units (MCU) with AD/DA converters (e.g., 8051 MCU) can be utilized to minimize cost.

4) Since the driving circuit is relatively independent of each other, the proposed gate drive has no restrictions on the switching duty cycle and the number of semiconductors in series-connection.

The rest of this article is organized as follows. Section II discusses the influence and the causes of gate discharge deviation on the drain-to-source voltage imbalance. In Section III, the structure and the operation principle of the proposed active gate drive are elaborated, followed by Section IV which presents the design guideline in detailed. The voltage balancing performance of the proposed gate drive is experimentally verified by a test board with two/three series-connected SiC MOSFETs in Section V. Finally, conclusions are drawn in Section VI.

## II. INFLUENCE AND CAUSES OF GATE DISCHARGE DEVIATION

### A. The Influence of Gate Discharge Deviation on Voltage Imbalance

The turn-off transient of SiC MOSFET can be seen as a discharge process of the input capacitance $C_{iss}$ which is constituted of the gate-source capacitance $C_{GS}$ and the gate-drain capacitance $C_{GD}$. Considering a SiC MOSFET driven by the voltage from positive ($V_{DD}$) to negative ($V_{EE}$), the electrical discharge $Q_{Goff}$ can be expressed as follows:

$$Q_{Goff} = Q_{GSoff} + Q_{GD} \quad (1)$$

where $Q_{GSoff}$ and $Q_{GD}$ are the discharges of $C_{GS}$ and $C_{GD}$, respectively. At the beginning of the turn-off period, when the drain-source voltage $V_{DS}$ has not started to rise, $Q_{GSoff}$ can be defined as [1] [15]:

$$Q_{GSoff} = (C_{GS} + C_{GDon})(V_{DD} - V_{miller}) \quad (2)$$

where $C_{GS}$ can be deemed as a constant due to its insignificant nonlinearity [15], and $C_{GDon}$ is the on-state value of $C_{GD}$. $V_{miller}$ is the miller stage voltage which has a relationship with the threshold voltage $V_{th}$, the drain current $I_D$ and the transconductance $g_m$ as [10]:

$$V_{miller} = V_{th} + \frac{I_D}{g_m} \quad (3)$$

From equation (2) and (3), it can be found that an increase in $I_D$ will cause a proportional increase in $V_{miller}$, which means

$Q_{GSoff}$ has a negative relationship with $I_D$. Therefore, for well-matched SiC MOSFETs in series-connection, the $Q_{GSoff}$ will be equal due to the same $I_D$.

The $Q_{GD}$ is required during the rise of $V_{DS}$. It can be calculated as:

$$Q_{GD} = \int_{V_{DSon}}^{V_{DSoff}} C_{GD(VDS)} dV_{DS} \quad (4)$$

where $V_{DSoff}$ and $V_{DSon}$ refer to the drain-source voltages in off-state and on-state, respectively. $C_{GD(VDS)}$ is the value of $C_{GD}$ varying dramatically with $V_{DS}$. In [1], the expression for calculating the value of $C_{GD(VDS)}$ is presented as:

$$C_{GD(VDS)} = K_G C_{ox} \bigg/ \sqrt{1 + \frac{2V_{DS}C_{ox}^2}{q\varepsilon_s N_D}} \quad (5)$$

From (4) and (5) we can obtain that

$$Q_{GD} = \frac{2K_G q\varepsilon_s N_D}{C_{ox}} \left( \sqrt{1 + \frac{2V_{DSoff}C_{ox}^2}{q\varepsilon_s N_D}} - \sqrt{1 + \frac{2V_{DSon}C_{ox}^2}{q\varepsilon_s N_D}} \right) \quad (6)$$

where $K_G$ is a geometry factor, $q$ is the fundamental electronic charge, $\varepsilon_s$ is the semiconductor dielectric constant, $N_D$ is the doping concentration of n-drift region, $C_{ox}$ is the specific capacitance of the gate oxide. By transforming the equation (6), we can obtain that

$$V_{DSoff} = \frac{q\varepsilon_s N_D}{2C_{ox}^2} \left[ \left( \frac{Q_{GD}C_{ox}}{2K_G q\varepsilon_s N_D} + \sqrt{1 + \frac{2V_{DSon}C_{ox}^2}{q\varepsilon_s N_D}} \right)^2 - 1 \right] \quad (7)$$

For series-connected SiC MOSFETs operated in high-voltage applications, the $V_{DSon}$ of each device is several orders of magnitude smaller than $V_{DSoff}$, and thus they can be considered equal. As can be seen form (7), there is a positive correlation between $Q_{GD}$ and $V_{DSoff}$.

With reference to two well-matched SiC MOSFETs in series-connection, if there is a gate discharge deviation $\Delta Q_{Goff}$ between them, by the analysis above, we can get the following conclusions:

1) The gate-drain discharge deviation $\Delta Q_{GD}$ will be reflected in $\Delta Q_{Goff}$ due to the nearly same $Q_{GSoff}$.

2) The $\Delta Q_{GD}$ plays a dominant role in the voltage distribution between the devices.

To demonstrate the effect of $\Delta Q_{Goff}$, a chopper circuit operating at the bus voltage $V_{DC}$ is established for simulation in LTspice. As shown in Fig. 1, $T_1$ and $T_2$ are the series-connecting SiC MOSFETs of the same model C2M0040120D from CREE. A parameter $\alpha$ which stands for the voltage imbalance ratio is defined as:

$$\alpha = \frac{|V_{DSoff1} - V_{DSoff2}|}{V_{DC}} \times 100\% \quad (8)$$

where $V_{DSoff1}$ and $V_{DSoff2}$ are the off-state drain-source voltages of $T_1$ and $T_2$, respectively. Fig. 2 shows the relationship between $\alpha$ and $\Delta Q_{Goff}$ based on the parameter scan simulations. $T_1$ and $T_2$ are driven by constant current sources of 100mA. $\Delta Q_{Goff}$ is realized by adding an external parallel pulse current source to the gate-source of $T_1$. From Fig. 2(a), with a constant $V_{DC}$ (1 kV), $\alpha$ is closely related to $\Delta Q_{Goff}$ but has little relationship with $I_D$ due to the same $Q_{GSoff}$ mentioned previously. From Fig. 2(b), with a constant $I_D$ (20A), $\alpha$ is not only positively correlated with $\Delta Q_{Goff}$ but also has a negative correlation with $V_{DC}$. This is mainly because a higher $V_{DC}$ indicates a higher $Q_{GD}$ required for turning off. For the same $\Delta Q_{GD}$, it means its proportion in $Q_{GD}$ is reduced, which suppresses its influence on voltage imbalance.

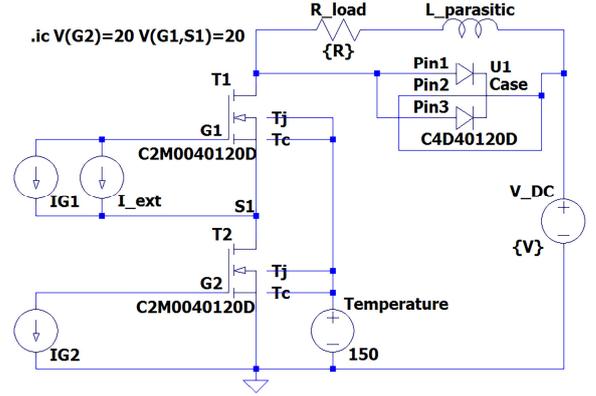

Fig. 1. Simulation model of series-connected SiC MOSFETs

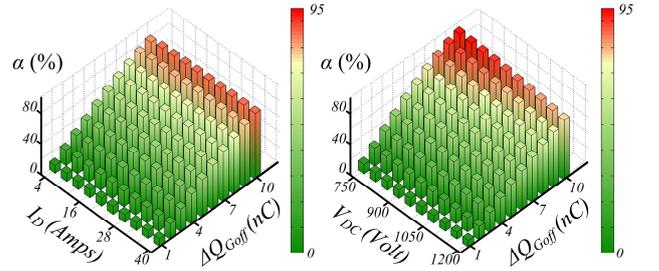

Fig. 2. Relationship between $\alpha$ (%) and $\Delta Q_{Goff}$ (nC). (a) $I_D$ changes from 4A to 40A while $V_{DC}$ is maintained at 1kV. (b) $V_{DC}$ changes from 750V to 1200V while $I_D$ is maintained at 20A.

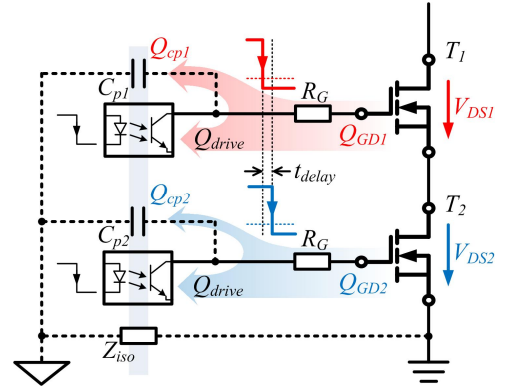

Fig. 3. Series-connected SiC MOSFETs with different gate-drain discharges.

*B. The Causes of Gate-Drain Discharge Deviation*

As concluded above, $\Delta Q_{GD}$ dominates the transient voltage imbalance ratio. Fig. 3 demonstrates the causes of $\Delta Q_{GD}$ between two intentionally matched SiC MOSFETs driven by conventional voltage drives. In the figure, $V_{DS}$ and $Q_{GD}$ with respective subscript number represent the drain-source voltage and the gate-drain discharge, respectively. $R_G$ is the external gate resistor of the same value. Fig. 4 shows the analytical waveforms during the turn-off transient, where $V_{GS1}$ and $V_{GS2}$ are the gate-source voltages of $T_1$ and $T_2$, respectively. From Fig. 4, the unbalanced voltage sharing is generated in the duration from $t_0$ to $t_5$, where the main causes of $\Delta Q_{GD}$ can be stated as follows:

1) Deviation between the propagation times of driving ICs: The turn-off propagation times of commercial gate driving ICs based on magnetic isolation (e.g., ADuM4135) or optical isolation (e.g., ACPL-W346) have inevitable deviations, which are from a dozen to dozens of nanoseconds between their minimum and maximum values. The gate discharge difference caused by the deviation time $t_{delay}$ is expressed as $\Delta Q_{delay}$, the value of which can be calculated by (9)

$$\Delta Q_{delay} = \frac{(V_{DD} - V_{miller})}{R_G} \cdot t_{delay} \quad (9)$$

where

$$t_{delay} = t_1 - t_0 = t_3 - t_2 \quad (10)$$

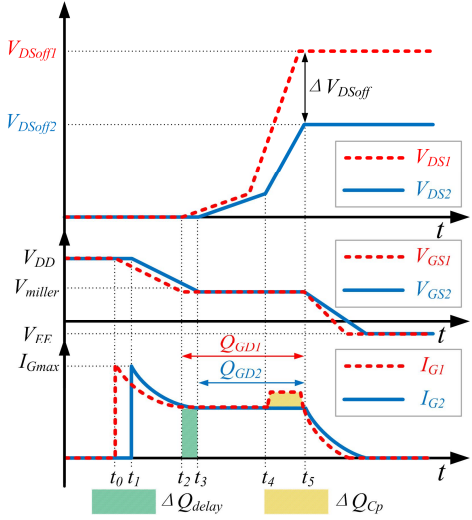

Fig. 4. $V_{DS}$, $V_{GS}$ and $I_G$ waveforms when there is a $\Delta Q_{GD}$ composed of $\Delta Q_{delay}$ and $\Delta Q_{Cp}$.

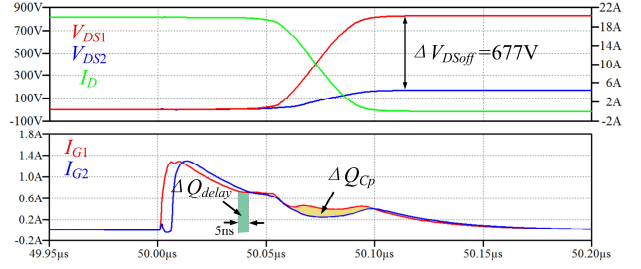

Fig. 5. Simulation results of $\Delta Q_{delay}$ and $\Delta Q_{Cp}$

TABLE I
THE PARAMETERS FOR SIMULATION

| Parameters | Values |
| --- | --- |
| SiC MOSFETs ($T_1$, $T_2$) | C2M0040120D |
| Drive voltages ($V_{DD}$, $V_{EE}$) | +20V, -5V |
| Parasitic capacitors ($C_{p1}$, $C_{p2}$) | 50pF |
| Gate resistor ($R_G$) | 15Ω |
| Driving time deviation ($t_{delay}$) | 5ns |

According to the waveforms of $I_G$ shown in Fig. 4, though $t_{delay}$ is generated from $t_0$ to $t_1$, the $\Delta Q_{delay}$ is located from $t_2$ to $t_3$ due to the same value of $Q_{GSoff}$. $\Delta Q_{delay}$ makes $V_{GS1}$ drop to $V_{miller}$ first, which means an earlier rising of $V_{DS1}$.

2) Different charges between the parasitic capacitors [26]: The turn-off period can be seen as a process of the gate drive extracting the electric charges stored in $C_{GD}$. As shown in Fig. 3, there is a parasitic capacitor $C_p$ which represents the total parasitic capacitances of the power and signal isolation barriers [28]. During the duration from $t_4$ to $t_5$, the dramatic increase of $V_{DS2}$ raises the potential of the source terminal of $T_1$ shifting up the voltage potential of $C_{p1}$ by nearly the same amount, which will extract more charge ($Q_{cp1}$) from $C_{GD}$. Compared to $Q_{cp1}$, the injected charge $Q_{cp2}$ is much smaller due to the insignificant variation of $V_{GS2}$. For the same gate drive operating under $V_{EE}$ during this period, the difference between $Q_{cp1}$ and $Q_{cp2}$ can be expressed as $\Delta Q_{cp}$, which can be calculated according to the following equation:

$$\begin{aligned}\Delta Q_{cp} &= C_{p1}\Delta V_{cp1} - C_{p2}\Delta V_{cp2} \\ &= C_{p1}\left[(V_{DS2} + V_{EE}) - (V_{DD} + R_{DS(on)}I_D)\right] \\ &\quad - C_{p2}(V_{EE} - V_{DD})\end{aligned} \quad (11)$$

where $R_{DS(on)}$ is the drain-source on-state resistance in milliohms, $\Delta V_{cp1}$ and $\Delta V_{cp2}$ represent the voltage variations of $C_{p1}$ and $C_{p2}$, respectively. Omitting the on-state voltage drop across the switch and assuming that $C_{p1}$ is equal to $C_{p2}$, equation (11) can be simplified to

$$\Delta Q_{cp} = C_{p1} \cdot V_{DS2} \quad (12)$$

The $\Delta Q_{Cp}$ will exist until $V_{DS1}$ and $V_{DS2}$ rise to their static values. Fig. 5 shows the simulated waveforms based on the circuit in Fig. 3 on the condition of $V_{DC}$=1 kV, $I_D$=20A. The parameters used for the simulation are listed in Table I. $\Delta Q_{delay}$ and $\Delta Q_{Cp}$ are pointed out to verify the above theoretical analysis, which led us to the conclusion that: 1) $\Delta Q_{delay}$ and $\Delta Q_{Cp}$, which make up $\Delta Q_{GD}$, have a great influence on the drain-source voltage balance; 2) the voltage imbalance can be alleviated by compensating $\Delta Q_{GD}$ precisely. Thereby, an active gate drive circuit aiming at compensating the deviation between gate-drain discharges of series-connected SiC MOSFETs will be presented in the next section.

III. PROPOSED ACTIVE GATE DRIVE

As mentioned in Section II, for series-connected SiC MOSFETs, $\Delta Q_{GD}$ will be reflected in $\Delta Q_{Goff}$ due to the approximately equal $Q_{GSoff}$. An active gate drive embedding closed-loop realization of $Q_{Goff}$ adjustment is presented in Fig. 6. It consists of three parts distinguished by different colors: 1) a current sink circuit providing additional gate-drain discharge to compensate $\Delta Q_{GD}$, 2) a trigger circuit to determine the time instants of additional gate discharge and $V_{DS}$ sampling, and 3) a sampling and control circuit used for taking $V_{DS}$ samples and controlling the amount of discharge in real-time. The structure and operating principle of each sub-circuit will be described in detail.

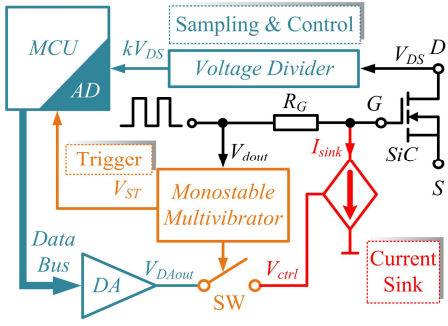

Fig. 6. Circuit diagram of the proposed active gate drive

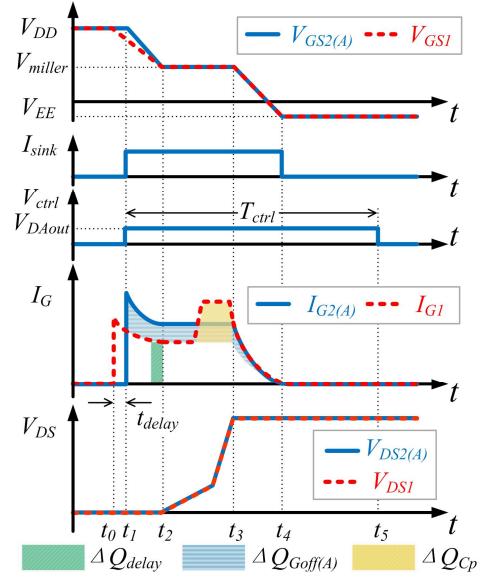

Fig. 8. Operation scheme of the proposed gate drive.

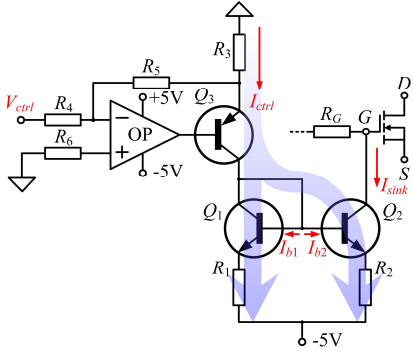

Fig. 7. Circuit implementation of the current sink circuit

*A. The Current Sink Circuit*

Fig. 7 shows the current mirror-based current sink circuit verified by [22], [29], and [30]. Fig. 8 shows the influenced turn-off waveforms of two series-connected SiC MOSFETs under its action. For comparison, we assume that the power circuit is the same as Fig. 3 except $T_2$ is driven by the proposed gate drive. In Fig. 8, the dotted lines stand for the waveforms of $T_1$ driven by conventional gate drive and the solid lines marked by a subscript *(A)* stand for the waveforms of $T_2$. $I_{sink}$ is the external gate current and the amount of discharge generated is written as $\Delta Q_{Goff(A)}$. $V_{ctrl}$ is the input control voltage of the current sink, value of which is zero or $V_{DAout}$ determined by the other two parts of the proposed drive circuit. The working process of $T_2$ affected by the current sink can be divided into four stages from $t_1$ to $t_5$.

*Stage $t_1 \sim t_2$*: The output voltage of the gate driving IC for $T_2$ drops from $V_{DD}$ to $V_{EE}$ at $t_1$ and the $C_{iss}$ of $T_2$ begins to discharge. In the meantime, $V_{ctrl}$ steps from zero to a positive value $V_{DAout}$. Consequently, $I_{sink}$ with a value calculated from

$$I_{sink} = I_{ctrl} = \frac{R_5}{R_3 R_4} \cdot V_{ctrl} \quad (13)$$

due to the same base currents of $Q_1$ and $Q_2$ ($I_{b1}=I_{b2}$), is generated to provide $\Delta Q_{Goff(A)}$. It can be seen from Fig. 8 that the required $Q_{GSoff}$ of $T_2$ is fulfilled faster due to a larger discharge current $I_{G2(A)}$.

*Stage $t_2 \sim t_3$*: At $t_2$, $V_{GS2(A)}$ drops to $V_{miller}$, almost all $I_{G2(A)}$ is used to discharge $C_{GD}$, and in the meantime, $V_{DS2(A)}$ starts to rise. According to the previous equation (7), if the amount of $\Delta Q_{Goff(A)}$ from $t_1$ to $t_3$ is equal to $\Delta Q_{GD}$, which is composed of $\Delta Q_{delay}$ and $\Delta Q_{Cp}$ from $t_1$ to $t_3$, then $V_{DSoff2}$ will be equal to $V_{DSoff1}$ and a balanced voltage sharing in off-state can be achieved.

*Stage $t_3 \sim t_4$*: At $t_3$, $V_{GS2(A)}$ and $V_{GS1}$ exit the miller plateau simultaneously. During this period, the falling speed of $V_{GS2(A)}$ from $V_{miller}$ to $V_{EE}$ is accelerated by the external gate discharge, which has little influence on the voltage sharing because both $V_{DS2(A)}$ and $V_{DS1}$ have arrived their static values.

*Stage $t_4 \sim t_5$*: At $t_4$, the turn-off process of $T_2$ has finished while the $V_{GS2(A)}$ is maintained as $V_{EE}$. Though the $V_{ctrl}$ is still at the value of $V_{DAout}$, the $I_{sink}$ is zero due to the almost equal voltage potentials of gate terminal and $V_{EE}$. During this stage, $I_{ctrl}$ will flow as the blue arrow shown in Fig. 7 and the bipolar junction transistor (BJT) $Q_2$ is deeply saturated because of the increased $I_{b2}$. In this period, $I_{ctrl}$ is no use but producing extra loss on the resistors and BJTs. Therefore, this stage should be as short as possible in practical application.

The LTSpice simulation results of series-connected SiC MOSFETs with the current sink are shown in Fig. 9. In order to demonstrate the effectiveness of the current sink, the simulation condition is the same as Table I and the component parameters of the current sink circuit are listed in Table II.

In comparison with Fig. 5, $\Delta Q_{Cp}$ shown in Fig. 9 becomes bigger due to the increased $V_{DS2}$ according to equation (12). However, due to the compensation effect of $\Delta Q_{Goff(A)}$, $\Delta V_{DSoff}$ shown in Fig. 9 is reduced to almost zero. From the waveforms of $V_{ctrl}$ and $I_{sink}$, it is worth mentioning that although $V_{ctrl}$ remains as 2.6 V within a predefined duration of 150 ns, $I_{sink}$ starts to fall as soon as $V_{DS2(A)}$ reaches its static value. It is also noteworthy that there exists a propagation delay time between $V_{ctrl}$ and $I_{sink}$ due to the propagation characteristics of the operational amplifier and BJTs. According to the zoomed-in waveforms shown in Fig. 10, the propagation delay time is measured as 16ns.

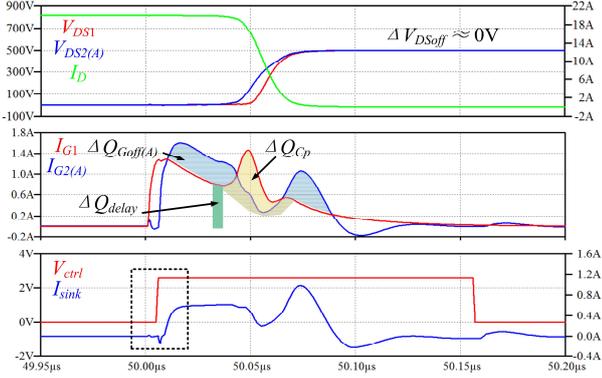

Fig. 9. Simulation results with the compensation discharge $\Delta Q_{Goff(A)}$.

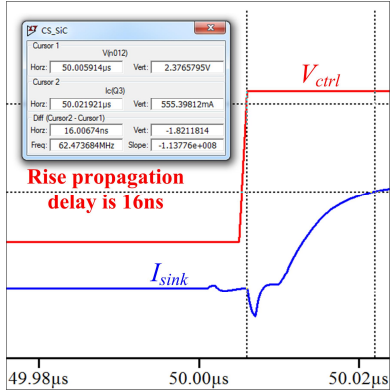

Fig. 10. Simulation results of rise propagation delay time between $V_{ctrl}$ and $I_{sink}$.

TABLE II
THE COMPONENT PARAMETERS OF THE CURRENT SINK CIRCUIT

| Components | Descriptions |
| --- | --- |
| Operational amplifier (OP) | LM7171 from Texas Instruments |
| NPN BJT ($Q_1$,$Q_2$) | ZXTN19060 from Diodes Inc. |
| PNP BJT ($Q_3$) | ZXTP19060 from Diodes Inc. |
| Resistors ($R_4$,$R_5$,$R_6$) | 1kΩ |
| Resistors ($R_3$) | 4Ω |
| Resistors ($R_1$,$R_2$) | 2Ω |
| Positive value of $V_{ctrl}$ ($V_{DAout}$) | 2.6V |

*B. The Trigger Circuit*

The trigger circuit is designed to determine the time interval during which the value of $V_{ctrl}$ is $V_{DAout}$ and the time instant when the sampling and control circuit starts analog-to-digital (AD) conversion. Fig. 11 shows the proposed trigger circuit. The +20/-5 V output voltage $V_{dout}$ from gate driving IC is scaled to +5/0 V by the voltage divider composed of resistors and capacitors. Then, the scaled-down gate signal is fed to two monostable multivibrator ICs (SN74LVC1G123 of Texas Instruments), which will produce a pulse of desired width at its falling edge, respectively. One pulse $V_{ST}$ with a width of $T_{ST}$ is sent to the sampling and control circuit to trigger the sampling on its falling edge. The $T_{ST}$ should be long enough to ensure that the $V_{DS}$ has reached $V_{DSoff}$ and remains stable. The other pulse with a width of $T_{ctrl}$ is used to connect $V_{ctrl}$ to the output voltage $V_{DAout}$ of the sampling and control circuit through an analog switch (TS5A63157 of Texas Instruments).

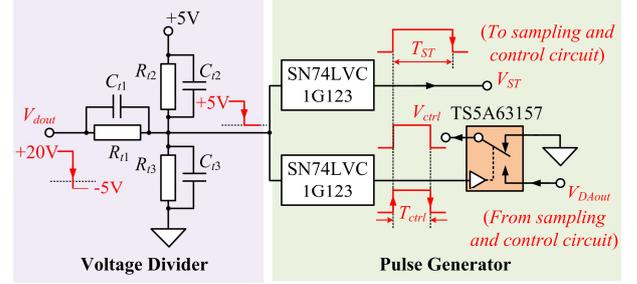

Fig. 11. Circuit implementation of the trigger circuit.

The response time of the trigger circuit for changing $V_{ctrl}$ should be as fast as possible to ensure high control precision and preventing the current sink circuit from producing too much extra loss. The voltage divider has no delay with approximate combination of resistors and capacitors. The maximum propagation delay time of SN74LVC1G123 at 5 V operating voltage is 7.6 ns and the maximum turn-on time of TS5A63157 for 5 V supply is 5 ns, so the maximum response time for the value change of $V_{ctrl}$ is 12.6 ns.

*C. The Sampling and Control Circuit*

Fig. 12 shows the detailed implementation of the sampling and control circuit. A compensated passive voltage divider is used to scale the high $V_{DS}$ to an acceptable AD input value, which is then buffered by a rail-to-rail amplifier GS8091 from Gainsil. At per switching cycle, a 12-bit AD converter (ADC) contained in a 1T 8051 core MCU (STC8A8K64S4A12 from STCmicro) starts the conversion as soon as it receives the external interrupt signal produced by the falling edge of $V_{ST}$. Two different kinds of regulators are built to process the measured $kV_{DS}$ ($k$ is the reduction factor of the voltage divider) into an output voltage $(n+1)^{th}V_o$ for the next switching cycle. First, a step-by-step regulator compares the error $e$ between the measured and the reference value with three threshold values ($e_{th1} \sim e_{th3}$) one by one, where the relationship among them satisfies

$$e_{th1} \gg e_{th2} \gg e_{th3} \qquad (14)$$

If $e$ is larger than $e_{thx}$ ($x=1 \sim 3$), a corresponding threshold $V_{thx}$ is added to the output voltage of the previous switching cycle ($n^{th}V_o$). Using the step-by-step regulator can ensure a fast determination of the output voltage range. Then, when $e$ is smaller than $e_{th3}$, a proportional-integral (PI) regulator is adopted to precisely adjust the $(n+1)^{th}V_o$ in a small range.

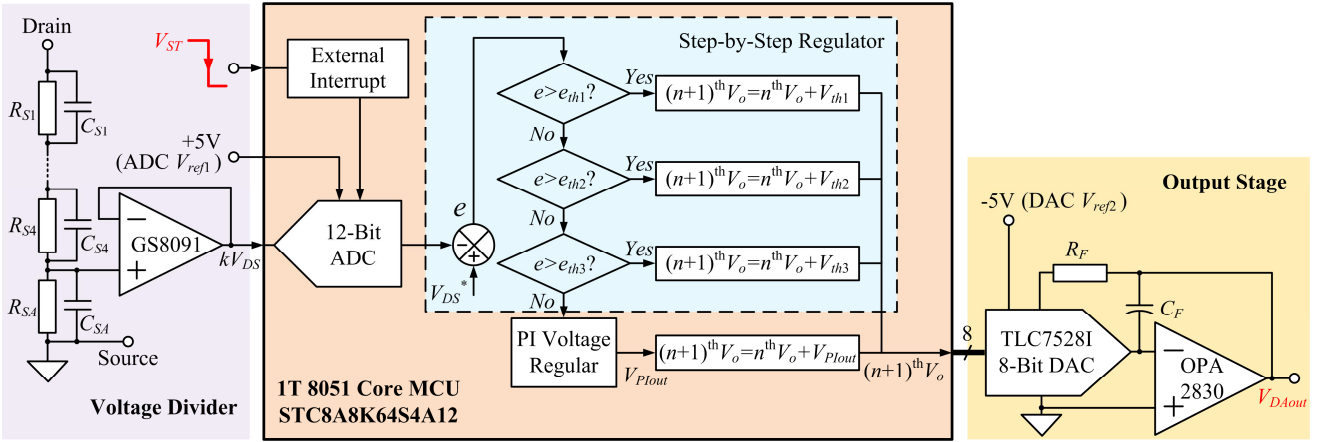

Fig. 12. Configuration of the sampling and control circuit.

Using the PI regulator instead of more and smaller stages of the step-by-step regulator can avoid a possible oscillation during the minimal step. The $(n+1)^{th}V_o$ is converted to an analog output voltage by an 8-bit digital-to-analog converter (DAC) and buffered by a rail-to-rail amplifier (TLC7528, OPA2830 from Texas Instruments) as shown in the output stage of Fig. 12, where $R_F$ and $C_F$ are used to prevent ringing or oscillation, $V_{DAout}$ is the final output voltage of the sampling and control circuit as aforementioned.

Although the turn-off transient of SiC MOSFET is ultra-fast, the calculation of $V_{DAout}$ can be done during the rest time of one switching period. The conversion rate of the ADC contained in STC8A8K64S4A12 is 800 KS/S, which means the sampling will be finished in 1.25 us. The running time of the algorithm in STC8A8K64S4A12 is about 3.6 us thanks to the 1T core of this MCU. The settling time and propagation delay time of TLC7528 are 100 ns and 80 ns, respectively. Therefore, the total delay time from the falling edge of $V_{ST}$ to the transition point of $V_{DAout}$ is approximately 5.03 us. Experimental verification of the actual delay time will be illustrated in Section V.

IV. DESIGN GUIDELINE FOR PROPOSED GATE DRIVE

Considering the DC chopper based on two series-connected SiC MOSFETs shown in Fig. 1, the design process of the primary parameters is under the rated condition, which is defined as: $V_{DC}$=1 kV, $I_D$=20 A, and the semiconductor device is C2M0040120D (1200 V/40 A) from CREE.

*A. Implementation of the Conventional Gate Drive*

According to the previous description of the active gate drive, there exist only three levels of the power supply voltages as +20 V for $V_{DD}$, +5 V and -5 V for all the auxiliary ICs and $V_{EE}$. An isolated DC/DC converter RKZ-052005D with the output voltage of +20 V/-5 V and a voltage regulator MC78L05 with +5 V output voltage are applied to provide the required voltage levels.

The gate drive IC is selected as ADuM4135 from Analog Devices and the external gate resistor is 15 Ω from YAGEO.

*B. Design of the Current Sink Circuit*

The current sink should have the capability of compensating the maximum $\Delta Q_{GD}$ composed of $\Delta Q_{delay}$ and $\Delta Q_{cp}$. $\Delta Q_{delay}$ can be estimated by equation (9). $V_{th}$ and $g_m$, which can be found in the datasheet of C2M0040120D, are 2.1 V and 13.2 S, respectively. So $V_{miller}$ is estimated to be 3.61 V. The difference between the minimum and maximum propagation delay times of ADuM4135 ($t_{delay}$) is 26 ns based on its datasheet. By substituting these values into equation (9), we can obtain the maximum value of $\Delta Q_{delay}$ is 28.4 nC. The maximum possible parasitic capacitances of RKZ-052005D and ADuM4135 are 135 pF and 2 pF, respectively. If the balanced voltage sharing is obtained, according to equation (12), the $\Delta Q_{cp}$ can be obtained as 68.5 nC. Therefore, the total maximum $\Delta Q_{GD}$ ($\Delta Q_{GD(max)}$) obtained by summing $\Delta Q_{delay}$ and $\Delta Q_{cp}$ is 96.9 nC.

In order to get wide output voltage range and fast response time, a high-speed op-amp MAX4213 with rail-to-rail output is selected as the input stage of the current sink. The PNP/NPN transistors are CPH6123/CPH6223 from ON semiconductor. These devices are different from the simulation models in previous section due to the lack of their Spice models in addition to their higher bandwidth and output swing. Table III and Table IV give the main characteristics of the op-amps and transistors used in experiment and simulation.

The Base-Emitter turn-on voltage ($V_{BE(on)}$) of CPH6123 is about 0.7 V, and the voltage drop on $R_3$ ($V_{R3}$) in Fig. 7 can be calculated by

$$V_{R3} = |V_{out-}| - V_{BE(on)} \quad (15)$$

where $V_{out-}$ is the negative peak value of the output swing. So the highest voltage drop on $R_3$ ($V_{R3(max)}$) is 3.8 V.

Fig. 13 shows the response time of the actual current sink circuit when $I_{sink}$ is 1 A, indicating a 21 ns delay time between $V_{ctrl}$ and $I_{sink}$, which is longer than the previous simulation result 16 ns due to the inevitable parasitic inductance in the loop. By taking account of the maximum response time 12.6ns of the trigger, the total maximum response time of the auxiliary circuit is 33.6 ns, theoretically. Considering the turn-off time $t_{off}$ of C2M0040120D with a 15 Ω gate resistor is approximately 125 ns by solving the equation:

$$t_{off} = t_{d(off)} + t_f \quad (16)$$

where $t_{d(off)}$ and $t_f$ are turn-off delay time and fall time from its "Switching Times vs. $R_{G(ext)}$" curve, respectively. Therefore, the minimum time remaining for discharge compensation ($t_{c(min)}$) is obtained as 91.4 ns. On the extreme condition that $\Delta Q_{GD(max)}$ needs to be compensated during $t_{c(min)}$, the value of $R_3$ can be solved by the following equation:

$$R_3 = \frac{V_{R3(max)}}{I_{ctrl(max)}} = \frac{t_{c(min)} V_{R3(max)}}{\Delta Q_{GD(max)}} \quad (17)$$

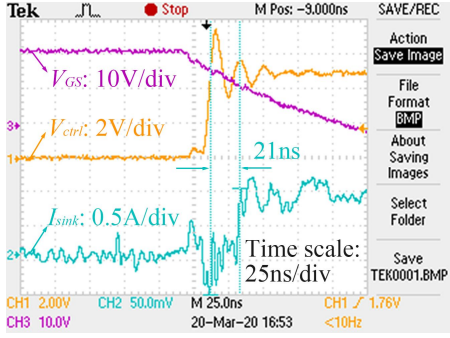

Fig. 13. Test waveforms for the response time of the current sink circuit. $I_{sink}$ is tested by a 0.1Ω sample resistor connected between the gate terminal of SiC MOSFET and the collector terminal of $Q_2$.

TABLE III
THE CHARACTERISTICS OF OP-AMPS

| Component | Values (±5V supply) | |
|---|---|---|
| | -3dB Bandwidth (BW) | Output Swing ($V_{out}$) |
| MAX4213 | 180MHz | -4.5V ~ +4.5V($R_L$=150Ω) |
| LM7171 | 140MHz | -2.8V ~ +2.8V($R_L$=100Ω) |

TABLE IV
THE CHARACTERISTICS OF TRANSISTORS

| Component | Gain-Bandwidth Product ($f_T$) |
|---|---|
| CPH6123 | 390MHz |
| CPH6223 | 380MHz |
| ZXTP19060 | 180MHz |
| ZXTN19060 | 130MHz |

By substituting the previously obtained parameters into (17), we can obtain the value of $R_3$ is approximately 3.58 Ω. In our case, $R_3$ is set as 3.6 Ω. To ensure the output capacity of the current sink, the maximum values of $R_1$ and $R_2$ ($R_{1(max)}$ and $R_{2(max)}$) can be calculated as

$$R_{1(max)} = R_{2(max)}$$
$$= \frac{|V_{EE}| - V_{R3(max)} - V_{CE3(sat)} - V_{CE1(sat)}}{I_{ctrl(max)}} \quad (18)$$

where $V_{CE3(sat)}$ and $V_{CE1(sat)}$ are the collector-to-emitter saturation voltages of $Q_3$ and $Q_1$, the maximum values of which are 230 mV and 130 mV, respectively. Therefore, $R_{1(max)}$ and $R_{2(max)}$ can be obtained as 0.77 Ω. In our case, two 0.5 Ω resistors in 2512 package are used as $R_1$ and $R_2$.

*C. Design of the Trigger Circuit*

The main design task of the trigger circuit is determining $T_{ST}$ and $T_{ctrl}$. As mentioned in Section III, when $I_{sink}$ is approximately zero and $Q_2$ is deeply saturated, $I_{ctrl}$ will generate extra loss on the transistors and resistors. From the perspective of power loss, the smaller $T_{ctrl}$ is the better. However, for the same $\Delta Q_{GD}$ to be compensated, too small $T_{ctrl}$ is corresponding to excessive $I_{ctrl}$, which will cause electromagnetic interference (EMI) issues and influence the gain-bandwidth product ($f_T$) of the BJTs according to the $f_T$-$I_C$ curve in their datasheets. Based on the above considerations and the minimum pulse width limitation of SN74LVC1G123, $T_{ctrl}$ is set as 125 ns, which is the same as the typical turn-off time of C2M0040120D with a 15 Ω gate resistor.

$T_{ST}$ is the time interval between the falling of $V_{GS}$ and the starting of the $V_{DS}$ sampling. On the one hand, $T_{ST}$ should be long enough to ensure that $V_{DS}$ has already been stable. On the other hand, $T_{ST}$ must be short enough to leave sufficient time for sampling. Equation (19) shows the $T_{ST}$ design constraint as

$$t_{off} < T_{ST} < T_{off(min)} - T_{sa(ADC)} \quad (19)$$

where $T_{off(min)}$ is the SiC MOSFET minimum off-state time, which depends on the switching frequency and duty cycle of the converter, and $T_{sa(ADC)}$ is the sampling time of ADC, which is 1.25 us as aforementioned. If the switching frequency is 40kHz, the duty cycle is limited in 10 % ~ 90 %, then $T_{ST}$ range can be determined: 125 ns < $T_{ST}$ < 1.25us. In our case, a 500 ns $T_{ST}$ is selected as a compromise.

*D. Design of the Sampling and Control Circuit*

The resistive part of the voltage divider is used for blocking the vast majority of $V_{DS}$, while acting as series-connected static balancing resistors. As shown in Fig. 12, it is substituted by four 100 kΩ ($R_{S1}$ ~ $R_{S4}$) resistors and a 2kΩ resistor ($R_{SA}$), corresponding to 4.97 V measured voltage on the condition that $V_{DS}$ is 1000 V, which is equal to the output swing of GS8091. The capacitive part is used for filtering out high-frequency oscillation and eliminating hysteresis quality caused by the parasitic inductances of $R_{S1}$ ~ $R_{S4}$ [26].

As discussed in Section II, for series-connected SiC MOSFETs, the relationship between α and $\Delta Q_{Goff}$ varies with a quite small proportion when $I_D$ changes. Therefore, the threshold values $e_{th1}$ ~ $e_{th3}$, $V_{th1}$ ~ $V_{th3}$, and PI coefficients (kp and ki) according to a fixed bus voltage have a certain representativeness and applicability. The selection of these values, which are listed in Table V, depends on the simulation results shown in Fig. 2.

The active gate drive board composed of the components mentioned above is shown in Fig. 14, where the sampling and control circuit is designed to be pluggable for replacement in the case of higher frequency MCU (e.g., FPGA) needed for ultra-high switching frequency. It can be seen from Fig. 14(a) that there exist no additional isolation stages for signals or power supplies, which makes the proposed gate drive superior

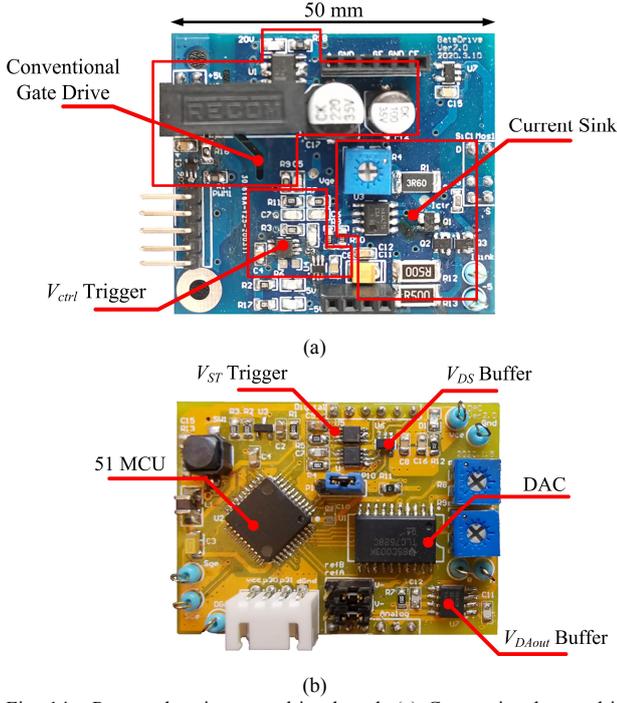

Fig. 14. Proposed active gate drive board: (a) Conventional gate drive with the current sink and the $V_{ctrl}$ trigger circuit. (b) Sampling and control circuit with the $V_{ST}$ trigger.

TABLE V
PARAMETERS OF THE CONTROL ALGORITHM

| Parameters | Values |
|---|---|
| $e_{th1}$, $e_{th2}$, $e_{th3}$ | 200V, 60V, 25V |
| $V_{th1}$, $V_{th2}$, $V_{th3}$ | 2V, 0.7V, 0.2V |
| $kp$, $ki$ | 0.01, 0.002 |

TABLE VI
COMPONENT PARAMETERS OF THE TEST BOARD

| Components | Values |
|---|---|
| Freewheeling diode ($D_F$) | WS3A006120E (1200V/9.5A) × 2<br>SMCJ400CA (400V/2.3A) × 4<br>33kΩ (2512) × 4 |
| Bus capacitor ($C_{Bus}$) | PFC 800V/10μF × 4 |
| Decoupling capacitor ($C_{dec}$) | MLCC 1000V/0.1μF × 10 |
| Bus balancing resistor ($R_{Bus}$) | 20kΩ (2512) × 10 |
| SiC MOSFETs ($T_1$, $T_2$) | C2M0040120D(1200V/40A) × 2 |

in terms of universality and cost effectiveness.

V. EXPERIMENTAL SETUP AND RESULTS

The active gate drive is experimentally verified in a test board configured as a DC chopper circuit. Fig. 15 shows the schematic and the photograph of the test board including two series-connected SiC MOSFETs. It can be observed that, the freewheeling diodes are composed of two series-connected

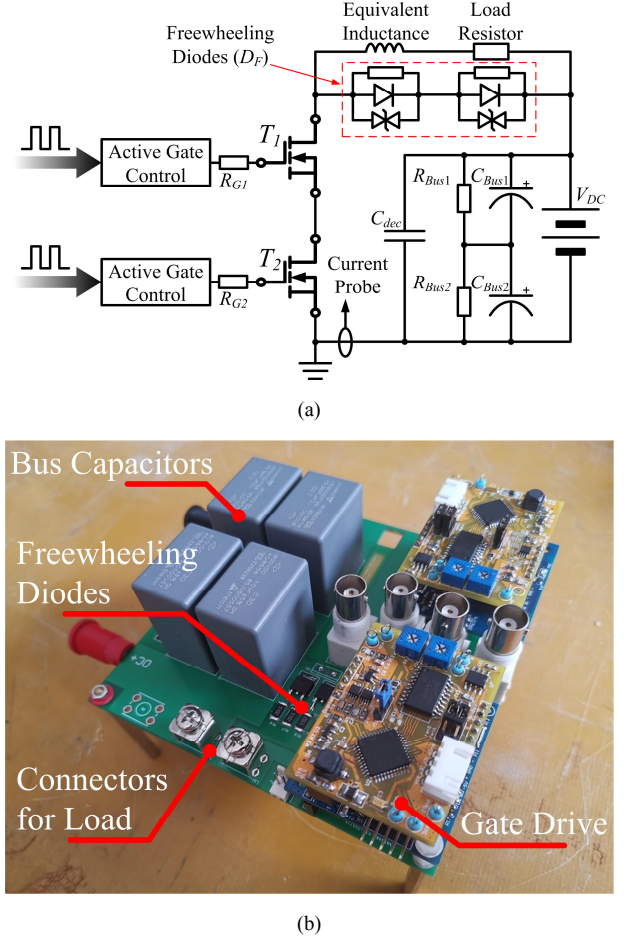

Fig. 15. Test board including two series-connected SiC MOSFETs with the proposed active gate drives. (a) Block schematic of the gate drive test board. (b) Hardware implementation of the test board (the SiC MOSFETs are mounted on the back).

SiC diodes with parallel-connected balancing resistors and TVS diodes, which are used for protecting SiC diodes from exceeding breakdown voltage caused by the overshoot of the transient unbalancing voltage [11]. The component parameters of the test board are listed in Table VI.

The test board operates at 1 kV bus voltage generated by a three-phase step-up transformer with a rectifier bridge and three series-connected bulk capacitors. The load is a 50Ω aluminum power resistor, the parasitic inductance of which is 30μH. A CPLD (5M240ZT100C5N from Altera) is utilized to generate driving signals, while the experimental waveforms are captured by a Tektronix TPS2024B (200 MHz, 2 GS/S, 4-isolated channels) digital oscilloscope with two 10X passive voltage probes (Tektronix P2220 200 MHz 300 V). Two 100X passive probes (UNI-T UT-P20 250 MHz 1.5 kV) are specifically used for measuring the $V_{DS}$. An AC current probe P6022 (Tektronix 120MHz) is used for capturing the waveforms of $I_D$ during the switching transient. The photograph of the experimental platform is shown in Fig. 16.

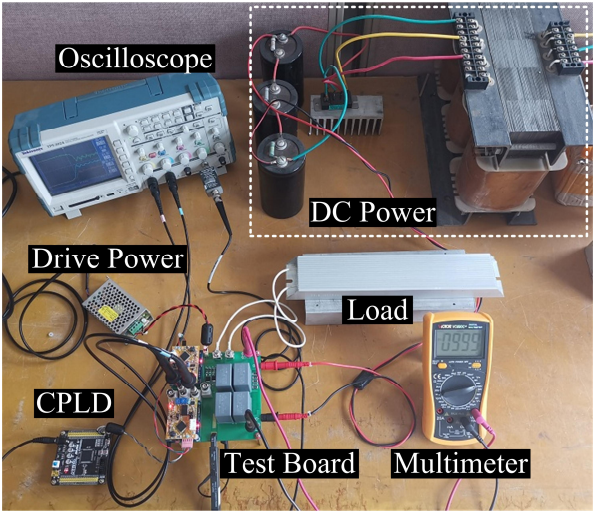

Fig. 16. Photograph of the experimental platform.

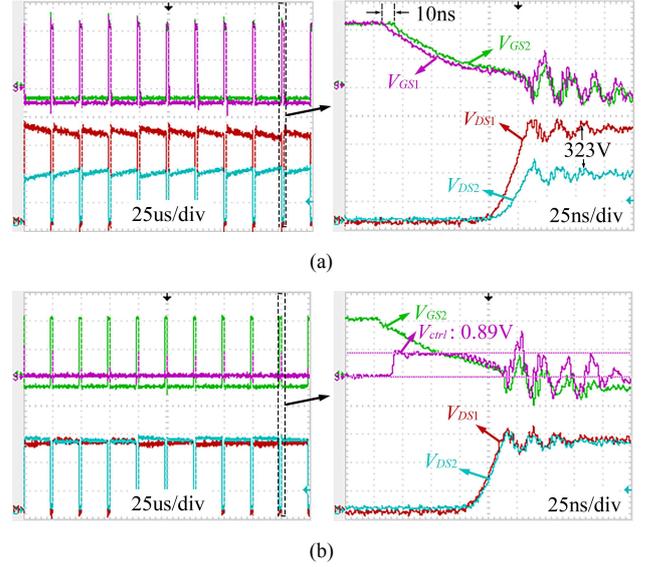

Fig. 17. Experimental waveforms when $t_{delay}$ is 10ns. Scale: $V_{DS}$ = 200V/div, $V_{GS}$ = 10V/div, $V_{ctrl}$ = 1V/div. (a) With only conventional gate drive. (b) With the proposed active gate drive.

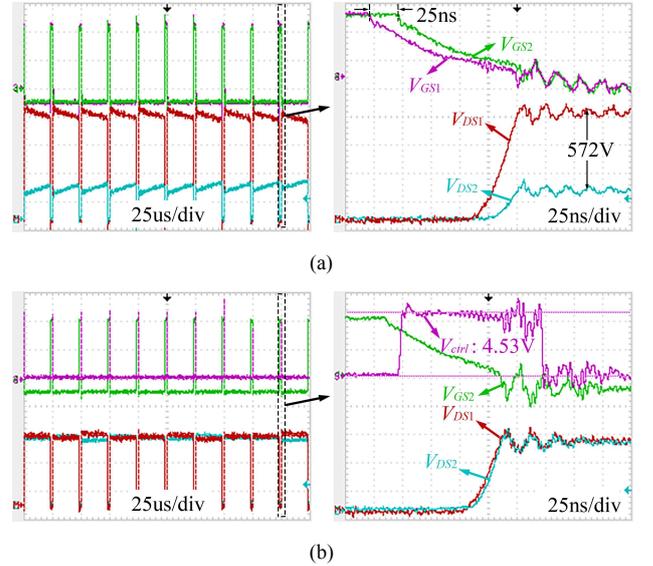

Fig. 18. Experimental waveforms when $t_{delay}$ is 25ns. Scale: $V_{DS}$ = 200V/div, $V_{GS}$ = 10V/div, $V_{ctrl}$ = 2V/div. (a) With only conventional gate drive. (b) With the proposed active gate drive.

*A. Verification of the Proposed Gate Drive Compensation Effect*

To validate the compensation effect of the proposed active gate drive on voltage balancing control, two comparison experiments are carried out at the switching frequency of 40kHz. The driving signals sent from CPLD are pre-adjusted to produce a fixed delay time $t_{delay}$ between the outputs of the driving ICs, so as to cause an intentional voltage imbalance. Fig. 17 shows the experimental waveforms of the two series-connected SiC MOSFETs with and without the proposed gate drives, respectively, where $t_{delay}$ is set as 10 ns. It should be noted that $V_{DS1}$ is displayed in the MATH channel by subtracting potentials of the drain terminals between $T_1$ and $T_2$. The oscillations in $V_{DS1}$ and $V_{DS2}$ waveforms are mainly caused by the resonance between the total equivalent output capacitance of the stacked devices and the power loop inductance [5]. Crosstalk is created between this $V_{DS}$ oscillations and the gate drive loop as well as the probe measurement loop, providing an EMI path of affecting both $V_{GS}$ and $V_{ctrl}$ dynamics in the form of ringing. With only the conventional gate drives, a voltage difference of 323 V between $V_{DSoff1}$ and $V_{DSoff2}$ is observed during the turn-off periods. This is mainly due to the $\Delta Q_{delay}$ generated by the preset $t_{delay}$ and the $\Delta Q_{CP}$ generated by different potentials between $C_{p1}$ and $C_{p2}$ as discussed in Section II. As shown in Fig. 17(b), $\Delta Q_{GD}$ is well compensated by a $V_{ctrl}$ with the amplitude of 0.89 V and a balanced voltage sharing is also obtained.

To further validate the performance of the proposed active gate drive, $t_{delay}$ is tuned to 25 ns which is almost the same as the time deviation between the minimum and the maximum propagation delay times of ADuM4135. As can be seen from the experimental results presented in Fig. 18(a), the voltage difference reaches 572 V during the turn-off transient due to the enlarged $\Delta Q_{delay}$. Waveforms in Fig. 18(b) present good balanced voltage distribution between the series-connected SiC MOSFETs driven by the proposed active gate drive, where the amplitude of $V_{ctrl}$ rises to 4.53 V to compensate the new $\Delta Q_{delay}$. Multi-pulse waveforms and their zoom-in views shown in Fig. 17 and 18 indicate that the proposed active gate drive has sufficient capability to compensate $\Delta Q_{GD}$ and achieve good dynamic and static voltage balancing.

*B. Influence of Switching Frequency and Duty Cycle*

From the design guideline of the proposed active gate drive mentioned in Section IV, the switching frequency ($f_s$) and the duty cycle are only considered when calculating the $T_{ST}$. Theoretically, as long as the switching period is longer than 5.5 us as illustrated in Section III_C, the control algorithm proposed in this paper can take effect. In order to verify the effectiveness of the proposed driving circuit at different switching frequencies, the experimental results with a 10ns $t_{delay}$ at half of the original switching frequency are shown in Fig. 19(a). It can be observed that a balancing voltage distribution can be achieved at 20 kHz. In contrast, the waveforms at a doubled switching frequency are also shown in Fig. 19(b). As can be observed, the dynamic and static voltage balance can still be achieved. It should be noted that $V_{ctrl}$ is maintained around 0.89V due to the same $\Delta Q_{delay}$ generated by the unchanged $t_{delay}$ according to equation (9).

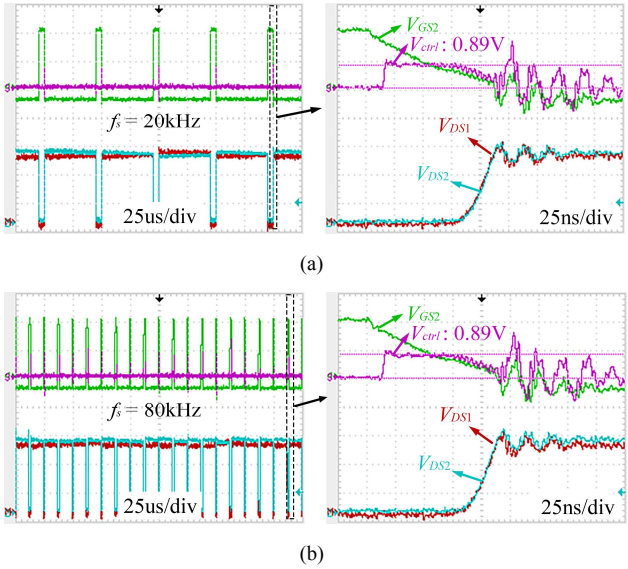

Fig. 19. Experimental waveforms under different switching frequencies ($t_{delay}$ = 10ns). Scale: $V_{DS}$ = 200V/div, $V_{GS}$ = 10V/div, $V_{ctrl}$ = 1V/div. (a) $f_s$ = 20 kHz. (b) $f_s$ = 80 kHz.

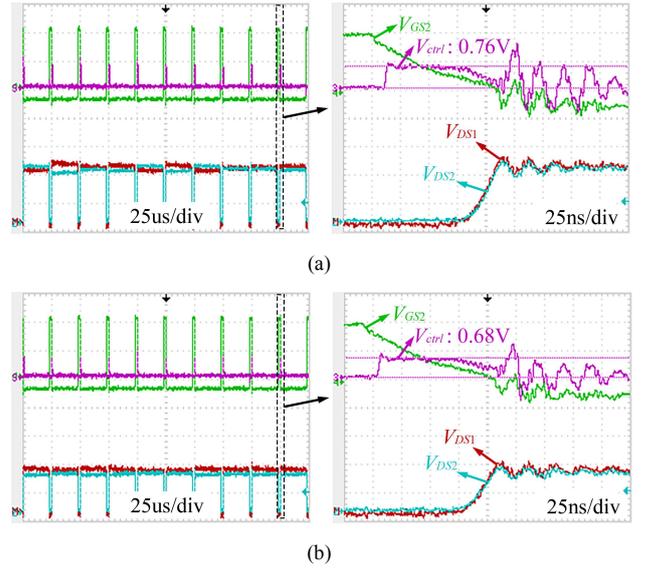

Fig. 21. Experimental waveforms under different DC-bus voltages ($t_{delay}$ = 10ns). Scale: $V_{DS}$ = 200V/div, $V_{GS}$ = 10V/div, $V_{ctrl}$ = 1V/div. (a) $V_{DC}$ = 800 V. (b) $V_{DC}$ = 600 V.

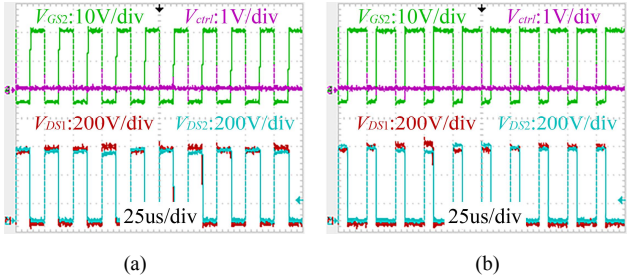

Fig. 20. Multi-pulse waveforms at different duty cycles ($t_{delay}$ = 10ns). (a) Duty cycle = 0.5. (b) Duty cycle = 0.7.

The proposed active gate drive is also tested at different duty cycles. Fig. 20 shows the multi-pulse waveforms at 0.5 and 0.7 duty cycles, respectively. From the evenly distributed voltages among the two devices, we can see that duty cycle is not a limitation on the voltage balance control of the proposed gate drive.

### C. Influence of DC-bus Voltage and External Gate Resistor

The influence of the DC-bus voltage is also discussed in this paper. Fig. 21 shows the experimental results under the condition that the DC-bus voltage is set as 800 V and 600 V, respectively, where the reference voltage $V_{DS}^*$ is set as 1.99 V and 1.49 V, respectively. The $t_{delay}$ is maintained as 10 ns. With other unchanged parameters in the control algorithm, a high $V_{DS}$ consistency can still be ensured. However, in comparison with Fig. 17(b), the amplitude of $V_{ctrl}$ is getting a little smaller as the decreasing of the DC-bus voltage. This is mainly attributed to the smaller $\Delta Q_{Cp}$ corresponding to lower $V_{DS2}$ according to equation (12).

The selection of the component parameters given in Section IV is based on the condition that $R_G$ = 15 Ω. According to the switching transient analysis given in [23] and [29], during the turn-off transient, $R_G$ plays a major role in the $t_{off}$ of SiC MOSFET and the peak value of $I_G$, which are the main basis in the parameter selection process. Fig. 22(a) illustrates the voltage balance effect of the proposed gate drive when $R_G$ is 12 Ω and $t_{delay}$ is 10 ns. In comparison with Fig. 17(b), for the same $\Delta Q_{GD}$ to be compensated, the balanced voltage distribution is achieved at the price of a much higher $V_{ctrl}$ (4.8V), which is equal to the maximum output of the sampling and control circuit. In Fig. 22(b), each $R_G$ is replaced by a 10Ω resistor, and the experimental results are presented. It can be seen that even with the maximum value of $V_{ctrl}$, a voltage difference cannot be eliminated with the currently selected components. This problem is mainly attributed to the reduced $t_{off}$, the value of which becomes 97 ns according to the "Switching Times vs. $R_{G(ext)}$" curve of C2M0040120D. It should be noted that the 85ns $t_{off}$ shown in Fig. 22(b) is an accelerated one. For a 10 ns $t_{delay}$, the $\Delta Q_{delay}$ is estimated as 17.14nC according to (3) and (9). The maximum $\Delta Q_{Cp}$ remains unchanged as 68.5nC. Thereby, the maximum $\Delta Q_{GD}$ that needs to be compensated is 85.64nC by adding $\Delta Q_{delay}$ and $\Delta Q_{Cp}$. However, due to the 33.6 ns response time of the auxiliary circuit mentioned in Section IV, the remaining time for the current sink is only 63.4ns which is corresponding to a 2.81 Ω $R_3$ according to (17). Therefore, a smaller $R_3$ seems to be a better choice when $R_G$ is reduced. Nevertheless, the resulting enlarged $I_{ctrl}$ will cause EMI issues and get the BJTs out of their active regions for fast response [31], which will shorten the remaining time again. Therefore, the selection of the transistors must be reconsidered, as well as other parts of the current sink.

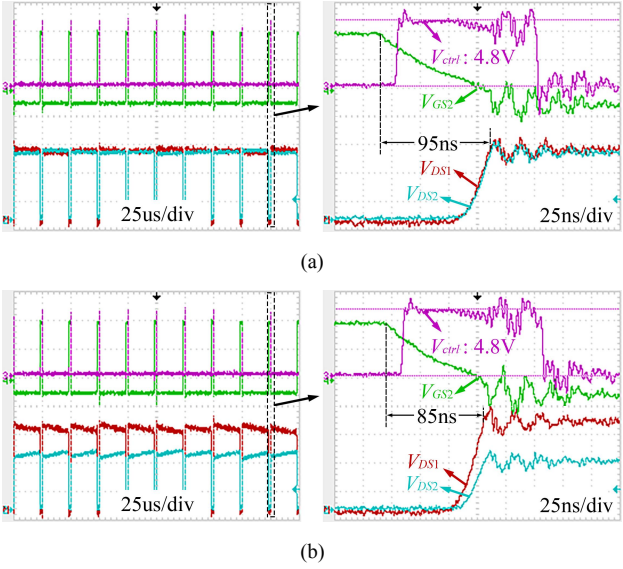

(a)

(b)

Fig. 22. Experimental waveforms with different external gate resistors ($t_{delay}$ = 10ns). Scale: $V_{DS}$ = 200V/div, $V_{GS}$ = 10V/div, $V_{ctrl}$ = 2V/div. (a) $R_G$ = 12 Ω. (b) $R_G$ = 10 Ω.

*D. Verification of the Voltage Balancing Algorithm under resistive load*

First of all, the time sequence among the relevant voltage signals of the sampling and control circuit is illustrated in Fig. 23. It can be observed that a positive pulse showing a preset period of 500 ns occurs on the $V_{ST}$ waveform as soon as $V_{GS2}$ begins to fall, during when $V_{DS2}$ has reached its stable value $V_{DS2off}$. The time interval between the falling edge of $V_{ST}$ and the rising edge of $V_{DAout}$ is measured as 5 us, which is consistent with the description in Section III. It consists of sampling time, algorithm running time and DAC output delay time.

Fig. 24 shows the waveforms of the incipient 18 switching cycles to verify the effectiveness of the control algorithm, where $t_{delay}$ is 25 ns. The reference value of $V_{DS}$ ($V_{DS}^*$) in the MCU is set as 2.49 V corresponding to the 500 V average voltage. It can be observed that there is a 2V rise in $V_{DAout}$ soon after the first rising edge of $V_{DS2}$ since that the error $e$ between $V_{DS}^*$ and $V_{DSoff2}$ is larger than $e_{th1}$. Then, in the next three cycles, the increment of $V_{DAout}$ drops to 0.7V and 0.2V due to the reduced $e$ which is lower than 100V. Until the fifth cycle, the voltage imbalance ratio gets balanced to ±5%, and then PI regulator takes place of the step-by-step regulator and stabilizes $V_{DAout}$ within a certain small range. The voltage balancing control is realized.

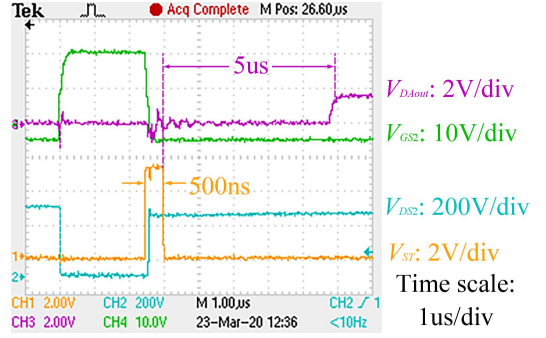

Fig. 23. Time sequence among the relevant voltage signals of the sampling and control circuit.

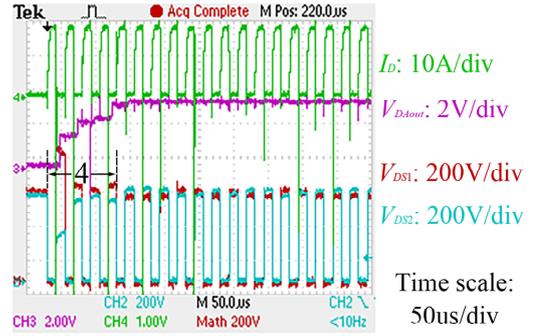

Fig. 24. The voltage regulating process using the active gate drive with the proposed control algorithm.

The proposed gate drive is also tested in the application of three devices in series-connection. Fig. 25 shows the corresponding experimental waveforms under 900V bus voltage, where CH1 to CH3 (100V/div) represent $V_{DS1}$ to $V_{DS3}$, respectively. As can be observed, it takes six switching cycles for the sampling and control circuits to adjust and identify the suitable output for each drive and maintain the voltage imbalance ratio in a small range.

*E. Verification of the Voltage Balancing Algorithm under inductive load*

As mentioned in Section II, the voltage imbalance ratio $\alpha$ is scarcely affected by $I_D$ for two identical SiC MOSFETs in series-connection. However, in practical applications, the characteristics of the stacked devices, especially their parasitic capacitances, are mismatched with each other. According to the voltage rising rate of a power MOSFET [33]:

$$\frac{dV_{DS}}{dt} = \frac{I_D + g_m \cdot (V_{th} - V_{EE})}{C_{GD} \cdot (1 + R_G \cdot g_m) + C_{DS}}, \quad (20)$$

for series-connected devices with different parasitic

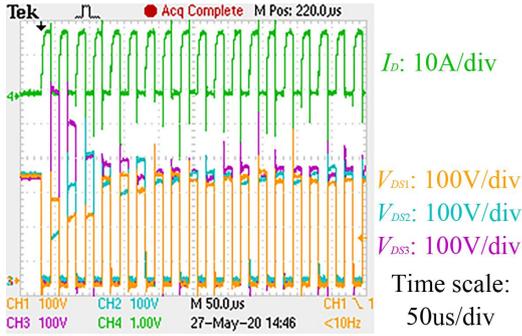

Fig. 25. The experimental results of three series-connecting SiC MOSFETs with the proposed active gate drive.

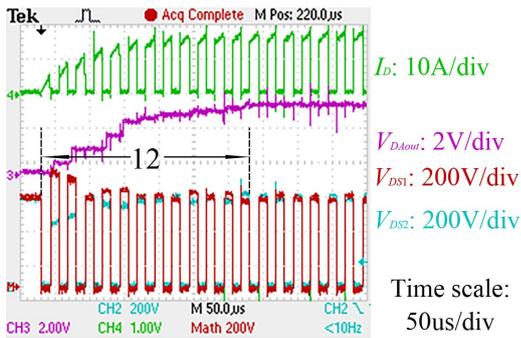

Fig. 26. Closed-loop test results under the resistive and inductive load.

TABLE VII
PROBES USED FOR THE SWITCHING LOSS EVALUATION

| Measured signals | $V_{DS}$ | $I_D$ |
|---|---|---|
| Probe model | UT-P20 | P6022 |
| Probe bandwidth | 200MHz[1] | 120MHz |
| Rise/fall time capability[2] | 8.75ns | 14.58ns |
| Estimated minimum rise/fall time | 14ns | 20ns |
| Measured minimum rise/fall time | 21ns | 18ns |

(1) The bandwidth of UT-P20 is 250MHz but is limited by the bandwidth of the oscilloscope (TPS2024: 200MHz).
(2) Rise/fall time is calculated based on [32] and [34] with 5 times margin.

*F. Switching Loss analysis*

According to the voltage regulating process of the proposed active gate drive, which is realized by accelerating the turn-off speed of the slower SiC MOSFET, the total turn-off loss ($E_{off\_total}$) of the series-connecting devices during the turn-off transient should be smaller. Moreover, since the active gate drive is not activated during the turn-on transient, the total turn-on loss ($E_{on\_loss}$) of the stacked devices will not be affected.

In order to verify the above inference, a double pulse test (DPT) is performed with and without the proposed active gate drive. According to [34], the switching loss evaluation of the wide band-gap device has a high requirement of the bandwidths and propagation delays of the probes and oscilloscope.

1) Bandwidth: Some literatures have discussed the relationship between the measurement accuracy and the bandwidth of the probe. Table VII lists the fasted rise time that can be measured accurately by the probes used in the DPT [32] [34]. The minimum rise/fall time estimated by the simulation results in Section III and the practical measured values are also given, which proves that the capabilities of the probes are sufficient.

2) Switching V-I timing alignment: The V-I timing misalignment caused by the deviation between the propagation delays of the voltage probe and the current probe is a critical factor in the accuracy of switching loss evaluation [34]. According to the methods proposed in [35] and [36], a 10Ω non-inductive resistance is used to replace the freewheeling diodes, and the load is removed. By simultaneously using different types of probes to measure the voltage and current of the resistance, the delay between the rising/falling edges of the measured voltage and current waveforms can be obtained. This delay will be taken into account during the data processing in MATLAB, thereby minimizing the evaluation deviation caused by the V-I timing misalignment.

The experimental results are shown in Fig. 27, where the switching energy loss is calculated by MATLAB based on the data exported from the oscilloscope. The $t_{delay}$ is set to 10ns. It can be observed that $E_{off\_total}$ is 303.4μJ when only the conventional gate drive is used. Under the condition of using the proposed driving circuit, the control algorithm is disabled, and $V_{ctrl}$ is preset to obtain a balanced voltage between the end of the first pulse and the beginning of the second pulse. As shown in Fig. 27(b), $E_{off\_total}$ is reduced to 257.0μJ, which is 84.7% of the previous.

capacitances, although the $I_D$ flowing through all devices are identical, the discrepancy between their parasitic capacitances ($C_{GD}$ or $C_{DS}$) will cause a $dV_{DS}/dt$ deviation. This deviation of voltage rising rate increases with $I_D$, which will cause a larger $\Delta V_{DSoff}$ during static periods. When an inductive load is connected into the DC chopper, it takes a number of switching cycles, which depends on the load inductor and the switching frequency, to reach the final value of $I_D$. This also means that $\Delta V_{DSoff}$ requires more switching cycles to reach its maximum value, which increases the time required for the voltage balancing control.

In order to demonstrate the performance of the proposed active gate drive under an inductive load, a 2.6mH air-core inductor in series with a 30Ω current-limiting resistor is selected as the load. The 25ns $t_{delay}$ still exists, and $f_s$ remains at 40kHz. The regulating process is shown in Fig. 26. It can be seen from Fig. 26 that in the first switching cycle, the $\Delta V_{DSoff}$ is not as large as shown in Fig. 24 due to a much smaller $I_D$. Therefore, the first step of $V_{DAout}$ rise is 0.7V instead of 2V. However, during the next several switching cycles, $\Delta V_{DSoff}$ does not show a significant decrease due to the increasing $I_D$. Until the sixth cycle when $I_D$ is stable, a suitable value of $V_{DAout}$ is gradually found by the sampling and control circuit to achieve good voltage balancing. As can be observed, the number of the required switching cycles is twelve, which is extended due to the load inductance.

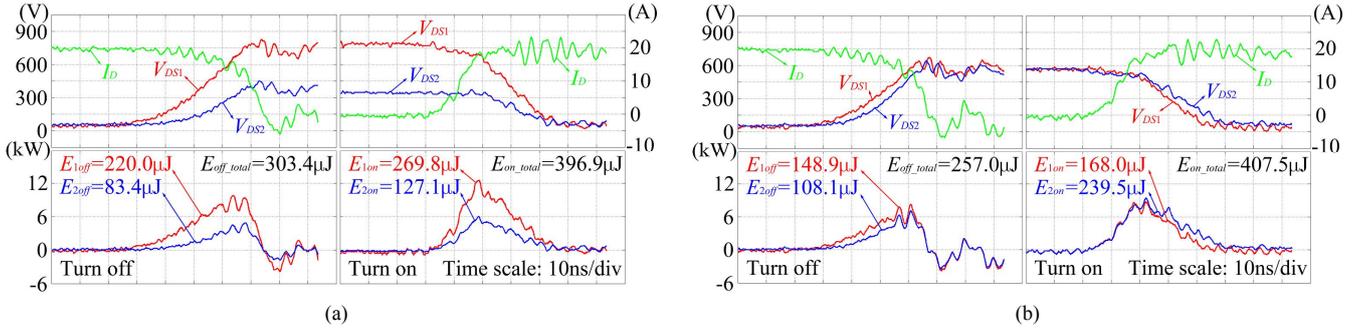

Fig. 27. The evaluated switching loss based on the results of the DPT when $t_{delay}$ is 10ns. (a) With only conventional gate drive. (b) With the proposed active gate drive.

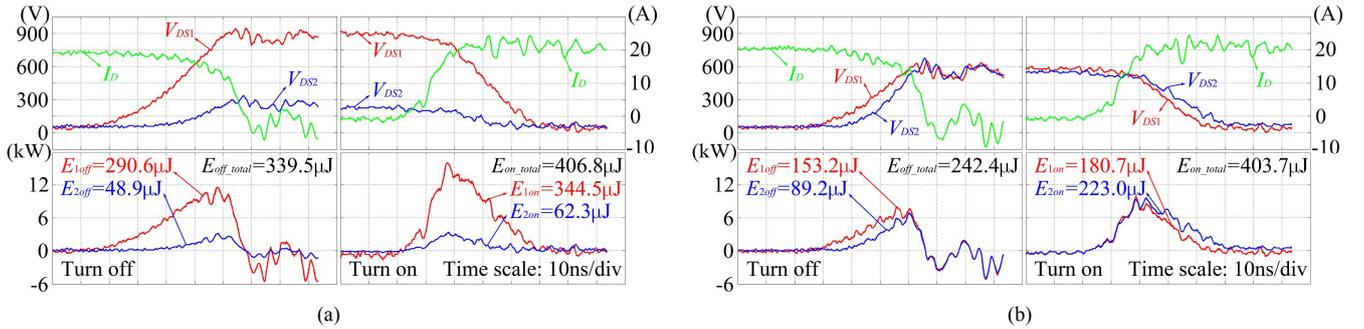

Fig. 28. The evaluated switching loss based on the results of the DPT when $t_{delay}$ is 25ns. (a) With only conventional gate drive. (b) With the proposed active gate drive.

Fig. 27 also shows the switching waveforms during the turn-on transient. It can be seen that no matter the voltage is balanced or not, $E_{on\_total}$ remains at around 400μJ. The switching loss is also measured when $t_{delay}$ is extended to 25ns, and $E_{off\_total}$ is reduced from 339.5μJ to 242.4μJ (71.4%) due to the increased turn-off speed of $T_2$ as shown in Fig. 28. From Fig. 27 and 28 we can get the conclusion that, when the auxiliary circuit is activated, the proposed active gate drive is able to reduce the total turn-off switching loss but scarcely effective on turn-on switching loss.

*G. Discussion*

Based on the above experimental results, in comparison with existing methods, the proposed method also has the following advantages while ensuring dynamic and static voltage balance:

1) Each active gate drive of the SiC MOSFETs in the stack is independent of each other, which allows an unlimited number of devices in series.

2) The auxiliary circuit is easy to be integrated with the commercial gate driving IC, only utilizing the falling edge of its output voltage as a trigger signal.

3) No additional isolations are required between the power side and the control side, which not only reduces costs but also avoids an increased $\Delta Q_{Cp}$ caused by the gate-to-ground parasitic capacitances in the isolation barriers [28][32].

4) The rates of sampling and control are equal to the switching frequency. Therefore, we can choose the most economical and suitable MCU and ADC/DAC based on the actual highest switching frequency of the converter, which provides flexibility for this method.

5) Since that the operational principle of the auxiliary circuit is to provide external $Q_{GD}$ for the slower device, so the switching transient of the entire series stack can be accelerated, which is beneficial for switching loss reduction.

According to the experimental results in Subsection V-C, future work can be focused on the following aspects:

1) The reference voltage for sampling and control circuit should be adjustable according to different DC-bus voltages. In order to realize it, either additional communication module or the conception of global and local controller mentioned in [22] shall be considered to employ.

2) With respect to a shorter turn-off time due to faster devices and smaller gate resistors, a tradeoff should be made between the $I_{sink}$ and the $f_T$ of the BJTs as aforementioned.

3) The control algorithm takes effect after the first falling edge of $V_{GS}$. Consequently, there exists at least one switching cycle when the auxiliary circuit does not work, which may lead to overvoltage issue and potential risk to system reliability. An energy recovery snubber circuit proposed and verified in [5] can be used to tackle the issue.

VI. CONCLUSION

This article presents a novel active gate drive that takes advantage of a controllable current sink to compensate the gate-drain discharge deviation between series-connecting SiC MOSFETs during turn-off transient. It can be realized by the proposed auxiliary circuit implemented on a commercial gate drive, using the original output of driving IC as a trigger signal. The auxiliary circuit achieves adaptive control through its built-in MCU, which avoids additional isolation barriers for signals and power supplies and makes a reduction in the costs and the gate-to-ground parasitic capacitance.

With a proper component selection and reasonable parameter setting, the proposed gate drive is tested by performing experiments on two and three series-connected SiC MOSFETs with a given driving signal deviation. As the experimental results have shown, even for different gate-drain discharges, switching frequencies and duty cycles, the proposed method does not need to change any parameters to obtain $V_{DS}$ curves that are almost coincident with each other. For different levels of DC-bus voltage, it is only necessary to reset the reference voltage $V_{DS}^*$ in the MCU to achieve a balanced voltage distribution. A simple method of real-time $V_{DS}^*$ tuning will be the direction of future research for this method.

REFERENCES

[1] B. J. Baliga, "Power MOSFETs," in *Fundamentals of Power Semiconductor Devices*. Cham, Switzerland: Springer, 2019, pp. 283-520.
[2] A. Marzoughi, A. Romero, R. Burgos and D. Boroyevich, "Comparing the state-of-the-art SiC MOSFETs: test results reveal characteristics of four major manufacturers' 900-V and 1.2-kV SiC devices," *IEEE Power Electron. Mag.*, vol. 4, no. 2, pp. 36-45, June 2017.
[3] A. Bolotnikov et al., "Overview of 1.2kV – 2.2kV SiC MOSFETs targeted for industrial power conversion applications," in *Proc. IEEE Appl. Power Electron. Conf. Expo. (APEC)*, Mar. 2015, pp. 2445-2452.
[4] T. Lu, Z. Zhao, H. Yu, S. Ji, L. Yuan and F. He, "Parameter design of a three-level converter based on series-connected HV-IGBTs," *IEEE Trans. Ind. Appl.*, vol. 50, no. 6, pp. 3943-3954, Nov.-Dec. 2014.
[5] F. Zhang, X. Yang, W. Chen and L. Wang, "Voltage balancing control of series-connected SiC MOSFETs by using energy recovery snubber circuits," *IEEE Trans. Power Electron.*, vol. 35, no. 10, pp. 10200-10212, Oct. 2020.
[6] T. C. Lim, B. W. Williams and S. J. Finney, "Active snubber energy recovery circuit for series-connected IGBTs," *IEEE Trans. Power Electron.*, vol. 26, no. 7, pp. 1879-1889, July 2011.
[7] M. Zarghani, S. Mohsenzade and S. Kaboli, "A series stacked IGBT switch based on a concentrated clamp mode snubber for pulsed power applications," *IEEE Trans. Power Electron.*, vol. 34, no. 10, pp. 9573-9584, Oct. 2019.
[8] R. Withanage and N. Shammas, "Series connection of insulated gate bipolar transistors (IGBTs)," *IEEE Trans. Power Electron.*, vol. 27, no. 4, pp. 2204-2212, April 2012.
[9] X. Wu, S. Cheng, Q. Xiao and K. Sheng, "A 3600 V/80 A series-parallel-connected silicon carbide MOSFETs module with a single external gate driver," *IEEE Trans. Power Electron.*, vol. 29, no. 5, pp. 2296-2306, May 2014.
[10] Y. Ren et al., "A compact gate control and voltage-balancing circuit for series-connected SiC MOSFETs and its application in a DC breaker," *IEEE Trans. Ind. Electron.*, vol. 64, no. 10, pp. 8299-8309, Oct. 2017.
[11] X. Song, A. Q. Huang, S. Sen, L. Zhang, P. Liu and X. Ni, "15-kV/40-A FREEDM supercascode: a cost-effective SiC high-voltage and high-frequency power switch," *IEEE Trans. Ind. Appl.*, vol. 53, no. 6, pp. 5715-5727, Nov.-Dec. 2017.
[12] A. Bagheri, H. Iman-Eini and S. Farhangi, "A gate driver circuit for series-connected IGBTs based on quasi-active gate control," *IEEE J. Emerg. Sel. Topics Power Electron.*, vol. 6, no. 2, pp. 791-799, June 2018.
[13] M. Shojaie, N. Elsayad, H. Moradisizkoohi and O. A. Mohammed, "Design and experimental verification of a high-voltage series-stacked GaN eHEMT module for electric vehicle applications," *IEEE Trans. Transport. Electrific.*, vol. 5, no. 1, pp. 31-47, March 2019.
[14] R. Wang, L. Liang, Y. Chen and Y. Kang, "A single voltage-balancing gate driver combined with limiting snubber circuits for series-connected SiC MOSFETs," *IEEE J. Emerg. Sel. Topics Power Electron.*, vol. 8, no. 1, pp. 465-474, March 2020.
[15] C. Yang et al., "A gate drive circuit and dynamic voltage balancing control method suitable for series-connected SiC MOSFETs," *IEEE Trans. Power Electron.*, vol. 35, no. 6, pp. 6625-6635, June 2020.
[16] T. Lu, Z. Zhao, S. Ji, H. Yu and L. Yuan, "Active clamping circuit with status feedback for series-connected HV-IGBTs," *IEEE Trans. Ind. Appl.*, vol. 50, no. 5, pp. 3579-3590, Sept.-Oct. 2014.
[17] S. Ji, T. Lu, Z. Zhao, H. Yu and L. Yuan, "Series-connected HV-IGBTs using active voltage balancing control with status feedback circuit," *IEEE Trans. Power Electron.*, vol. 30, no. 8, pp. 4165-4174, Aug. 2015.
[18] S. Ji, F. Wang, L. M. Tolbert, T. Lu, Z. Zhao and H. Yu, "An FPGA-based voltage balancing control for multi-HV-IGBTs in series connection," *IEEE Trans. Ind. Appl.*, vol. 54, no. 5, pp. 4640-4649, Sept.-Oct. 2018.
[19] K. Wada and K. Shingu, "Voltage balancing control for series connected MOSFETs based on time delay adjustment under start-up and steady-state operations," in *Proc. IEEE Energy Convers. Congr. Expo. (ECCE)*, Sep. 2018, pp. 5495-5499.
[20] T. Wang, H. Lin and S. Liu, "An active voltage balancing control based on adjusting driving signal time delay for series-connected SiC MOSFETs," *IEEE J. Emerg. Sel. Topics Power Electron.*, vol. 8, no. 1, pp. 454-464, March 2020.
[21] P. R. Palmer, J. Zhang and X. Zhang, "SiC MOSFETs connected in series with active voltage control," in *Proc. IEEE 3rd Workshop Wide Bandgap Power Devices Appl. (WiPDA)*, Nov. 2015, pp. 60-65.
[22] F. Zhang, X. Yang, Y. Ren, L. Feng, W. Chen and Y. Pei, "A hybrid active gate drive for switching loss reduction and voltage balancing of series-connected IGBTs," *IEEE Trans. Power Electron.*, vol. 32, no. 10, pp. 7469-7481, Oct. 2017.
[23] I. Baraia, J. A. Barrena, G. Abad, J. M. Canales Segade and U. Iraola, "An experimentally verified active gate control method for the series connection of IGBT/diodes," *IEEE Trans. Power Electron.*, vol. 27, no. 2, pp. 1025-1038, Feb. 2012.
[24] A. Raciti, G. Belverde, A. Galluzzo, G. Greco, M. Melito and S. Musumeci, "Control of the switching transients of IGBT series strings by high-performance drive units," *IEEE Trans. Ind. Electron.*, vol. 48, no. 3, pp. 482-490, June 2001.
[25] G. Belverde, A. Galluzzo, M. Melito, S. Musumeci and A. Raciti, "Snubberless voltage sharing of series-connected insulated-gate devices by a novel gate control strategy," *IEEE Trans. Power Electron.*, vol. 16, no. 1, pp. 132-141, Jan. 2001.
[26] A. Marzoughi, R. Burgos and D. Boroyevich, "Active gate-driver with dv/dt controller for dynamic voltage balancing in series-connected SiC MOSFETs," *IEEE Trans. Ind. Electron.*, vol. 66, no. 4, pp. 2488-2498, April 2019.
[27] Z. Zhang et al., "SiC MOSFETs gate driver with minimum propagation delay time and auxiliary power supply with wide input voltage range for high-temperature applications," *IEEE J. Emerg. Sel. Topics Power Electron.*, vol. 8, no. 1, pp. 417-428, March 2020.
[28] V. Nguyen, L. Kerachev, P. Lefranc and J. Crebier, "Characterization and analysis of an innovative gate driver and power supplies architecture for HF power devices with high dv/dt," *IEEE Trans. Power Electron.*, vol. 32, no. 8, pp. 6079-6090, Aug. 2017.
[29] F. Zhang, X. Yang, Y. Ren, L. Feng, W. Chen and Y. Pei, "Advanced active gate drive for switching performance improvement and overvoltage protection of high-power IGBTs," *IEEE Trans. Power Electron.*, vol. 33, no. 5, pp. 3802-3815, May 2018.
[30] H. Niu and R. D. Lorenz, "Real-time junction temperature sensing for silicon carbide MOSFET with different gate drive topologies and different operating conditions," *IEEE Trans. Power Electron.*, vol. 33, no. 4, pp. 3424-3440, April 2018.
[31] Shihong Park and T. M. Jahns, "Flexible dv/dt and di/dt control method for insulated gate power switches," *IEEE Trans. Ind. Appl.*, vol. 39, no. 3, pp. 657-664, May-June 2003.
[32] C. Liu, Z. Zhang, Y. Liu, Y. Si and Q. Lei, "Smart self-driving

multilevel gate driver for fast switching and crosstalk suppression of SiC MOSFETs," *IEEE J. Emerg. Sel. Topics Power Electron.*, vol. 8, no. 1, pp. 442-453, March 2020.
[33] P. Jeannin, D. Frey, and J. Schanen, "Sizing method of external capacitors for series association of insulated gate components," in *Proc. Eur. Power Electron. Drives Assoc. Conf.,* Graz, Austria, 2001.
[34] Z. Zhang, B. Guo, F. Wang, E. A. Jones, L. M. Tolbert and B. J. Blalock, "Methodology for wide band-gap device dynamic characterization," *IEEE Trans. Power Electron.*, vol. 32, no. 12, pp. 9307-9318, Dec. 2017.
[35] B. Callanan. (2011). *SiC MOSFET Double Pulse Fixture*. [Online]. Available: http://www.cree.com
[36] E. Jones et al., "Characterization of an enhancement-mode 650-V GaN HFET," in *Proc. IEEE Energy Convers. Congr. Expo.*, Sep. 2015, pp. 400–407.

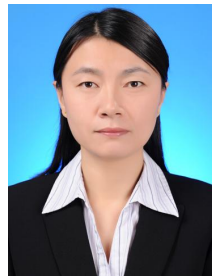

**Dan Yang** received the B.S. and M.S. degrees in biomedical engineering from Northeastern University (NEU), Shenyang, China, in 2002 and 2005, respectively, and the Ph.D. degree in detection and automatic control engineering from NEU, 2009.

From 2009 to 2012, she was a Postdoctoral researcher with the Computer and Science Engineering in NEU. From 2013 July to 2014 July, she was a visiting scholar sponsored by China Scholarship Council, with the School of Electrical and Computer Engineering, Oklahoma State University, Stillwater, USA. Now, she is an associate professor in Key Laboratory of Infrared Optoelectric Materials and Micro-Nano Devices at NEU, Shenyang, China.

Her current research includes wearable computing, physiological signal detection, and biological electromagnetic systems.

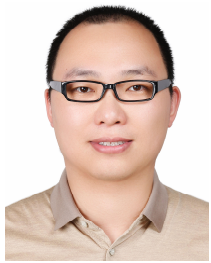

**Ye Zhou** received the B.S. and M.S. degrees in electrical engineering from Southwest Jiaotong University, Chengdu, China, in 2008 and 2011, respectively. He is currently pursuing the Ph.D. degree with the School of Information Science and Engineering, Northeastern University, Shenyang, China.

His research interests include applications of wide bandgap power semiconductor devices, as well as high power conversion system.

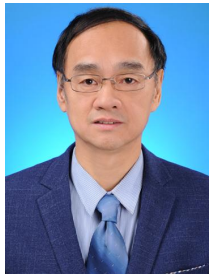

**Xu Wang** received the B.S. and M.S. degrees in industrial automation from the Northeastern University of Technology, Shenyang, China, in 1982 and 1986, respectively, and the Ph.D. degree in neurobiology from the Darmstadt University of Technology, Germany, in 1992.

From 1993 to 1998, he was with the National Institute of Neurobiology, Germany, as a Scientist and Post-Doctoral. Since 1998, he has been with the Northeastern University, Shenyang, China, where he became a Full Professor and a Doctoral Supervisor.

His research interests include power electronics, biology information, intelligent control, and neural network.

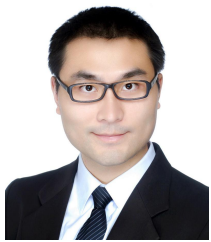

**Liang Xian** (S'11 – M'16) received B.S. and M.S. degrees in electrical engineering and power electronics from Southwest Jiaotong University, Chengdu, China, in 2008 and 2011, respectively, and Ph.D. degree in power engineering from Nanyang Technological University (NTU), Singapore in 2016.

From 2016 to 2019, he was a Scientist with Experimental Power Grid Centre (EPGC), Agency for Science, Technology and Research (A*STAR), Singapore. Currently he is a Senior Research Scientist with Energy Research Institute at NTU (ERI@N), Singapore.

His research interests include high power conversion system, distributed energy resource integration and control, wide bandgap semiconductor devices, and more-electric-aircraft power system.